\documentclass[showpacs,amssymb,twocolumn,superscriptaddress,prl]{revtex4-1}

\usepackage{amsmath,textcomp,gensymb}
\usepackage[dvips]{graphicx}

\begin{document}


\title{Damping of Nanomechanical Resonators}


\author{Quirin P. Unterreithmeier}
\email[]{quirin.unterreithmeier@physik.uni-muenchen.de}
\author{Thomas Faust}
\author{J\"org P. Kotthaus}
\affiliation{Fakult{\"a}t f{\"u}r Physik and Center for NanoScience (CeNS), Ludwig-Maximilians-Universit{\"a}t, Geschwister-Scholl-Platz 1, D-80539 M{\"u}nchen, Germany}


\date{\today}

\begin{abstract}
We study the transverse oscillatory modes of nanomechanical silicon nitride strings under high tensile stress as a function of geometry and mode index $m \leq 9$. Reproducing all observed resonance frequencies with classical elastic theory we extract the relevant elastic constants. Based on the oscillatory local strain we successfully predict the observed mode-dependent damping with a single frequency independent fit parameter. Our model clarifies the role of tensile stress on damping and hints at the underlying microscopic mechanisms.
\end{abstract}

\maketitle
The resonant motion of nanoelectromechanical systems receives a lot of recent attention. Their large frequencies, low damping i.e. high mechanical quality factors, and small masses make them equally important as sensors\,\cite{Jensen2008,Lassagne2008,LaHaye2009,Teufel2009} and for fundamental studies\,\cite{Aldridge2005,LaHaye2009,Rocheleau2010,Li2008,Teufel2009,Etaki2008,Li2008,Eichenfield2009}. In either case, low damping of the resonant motion is very desirable. Despite significant experimental progress\,\cite{Verbridge2006,Huttel2009}, a satisfactory understanding of the microscopic causes of damping is not yet achieved. Here we present a systematic study of the damping of doubly-clamped resonators fabricated out of prestressed silicon nitride leading to high mechanical quality factors\,\cite{Verbridge2006}. Reproducing the observed mode frequencies applying continuum mechanics, we are able to quantitatively model their quality factors by assuming that damping is caused by the local strain induced by the resonator's displacement. Considering various microscopic mechanisms, we conclude that the observed damping is most likely dominated by dissipation via localized defects uniformly distributed throughout the resonator.

We study the oscillatory response of nanomechanical beams fabricated from high stress silicon nitride (SiN). A released doubly-clamped beam of such a material is therefore under high tensile stress, which leads to high mechanical stability\,\cite{Unterreithmeier2009} and high mechanical quality factors\,\cite{Verbridge2006}. Such resonators are therefore widely used in recent experiments\,\cite{Eichenfield2009,Rocheleau2010}.
Our sample material consists of a silicon substrate covered with 400\,nm thick silicon dioxide serving as sacrificial layer and a $h=100$\,nm thick SiN device layer. Using standard electron beam lithography and a sequence of reactive ion etch and wet-etch steps, we fabricate a series of resonators having lengths of $35/n\,\micro{\rm m}, n=\{1,\dots,7\}$ and a cross section of $100\cdot 200$\,nm as displayed in Fig.\,1a and b. Since the respective resonance frequency is dominated by the large tensile stress\,\cite{Verbridge2006,Unterreithmeier2009a}, this configuration leads to resonances of the fundamental modes that are approximately equally spaced in frequency. Suitably biased gold electrodes processed beneath the released SiN strings actuate the resonators via dielectric gradient forces to perform out-of-plane oscillations, as explained in greater detail elsewhere~\cite{Unterreithmeier2009}. The length and location of the gold electrodes is properly chosen to be able to also excite several higher order modes of the beams. The experiment is carried out at room temperature in a vacuum below $10^{-3}$\,mbar to avoid gas friction.

\begin{figure}
  \includegraphics{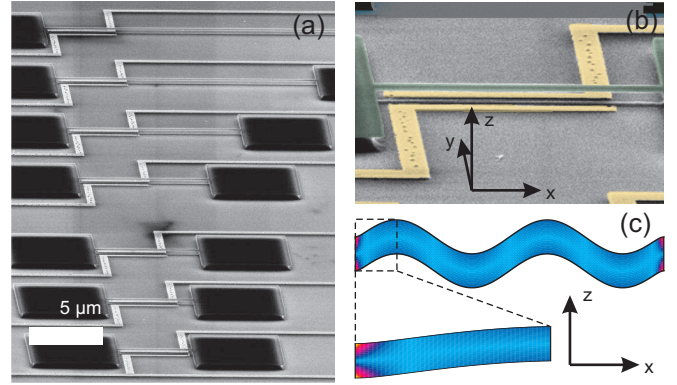}
  \caption{\label{fig1} {\bf Setup and Geometry}
  {\bf a} SEM-picture of our sample; the lengths of the investigated nanomechanical silicon nitride strings are $35/n\,\micro{\rm m}, n=\{1,\dots,7\}$; their widths and heights are 200\,nm and 100\,nm, respectively. {\bf b} Zoom-in of {\bf a}: the resonator (highlighted in green) is dielectrically actuated by the nearby gold electrodes (yellow); its displacement is recorded with an interferometric setup. {\bf c} Mode profile and absolute value of the resulting strain distribution (color-coded) of the longest beam's 4th harmonic as calculated by elastic theory.
  }
\end{figure}

The displacement is measured using an interferometric setup that records the oscillatory component of the reflected light intensity with a photodetector and network analyser\,\cite{Unterreithmeier2009,Azak2007}. The measured mechanical response around each resonance can be fitted using a Lorentzian lineshape as exemplarily seen in the inset of Fig.\,2. The thereby obtained values for the resonance frequency $f$ and quality factor $Q$ for all studied resonators and observed modes are displayed in Fig.\,2 (filled circles). In order to reproduce the measured frequency spectrum, we apply standard beam theory (see e.g.\,\cite{Timoshenko}). Without damping, the differential equation describing the spatial dependence of the displacement for a specific mode $m$ of beam $n$ $u_{\rm n,m}[x]$ at frequency $f_{\rm n,m}$ writes (with $\rho = 2800 {\rm kg/m}^3$ being the material density\,\cite{MEMShandbook}; $E_1$, $\sigma_0$ are the (unknown) real Young's modulus and built in stress, respectively)

\begin{equation}
\frac{1}{12} E_1 h^2 \frac{\partial^4}{\partial x^4} u_{\rm n,m}[x] - \sigma_0 \frac{\partial^2}{\partial x^2} u_{\rm n,m}[x] - \rho (2 \pi f_{\rm n,m})^2  u_{\rm n,m}[x]=0
\end{equation}

Solutions of this equation have to satisfy the boundary conditions of a doubly-clamped beam (displacement and its slope vanish at the supports ($u_{\rm n,m}[\pm l/(2n)] = (\partial/\partial x) u_{\rm n,m}[\pm l/(2n)] = 0$, $l/n$: beam length). These conditions lead to a transcendental equation that is numerically solved to obtain the frequencies $f_{\rm n,m}$ of the different modes.

The results are fitted to excellently reproduce the measured frequencies, as seen in Fig.\,2 (hollow squares). One thereby obtains as fit parameters the elastic constants of the micro-processed material $E_1=160\,$GPa, $\sigma_0= 830$\,MPa, in good agreement with previously published measurements~\cite{Unterreithmeier2009a}.

\begin{figure}
  \includegraphics{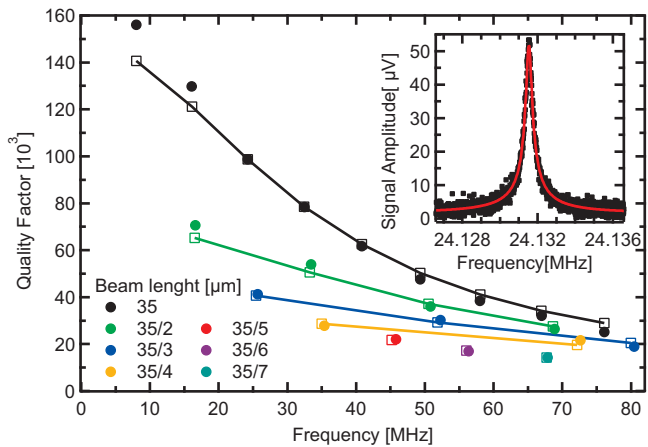}
  \caption{\label{fig2} {\bf Resonance Frequency and Mechanical Quality Factor}
  The harmonics of the nanomechanical resonator show a Lorentzian response (exemplary in the inset). Fitting yields the respective frequency and mechanical quality factor. The main figure displays these values for several harmonics (same color) of different beams as indicted by the color. To reproduce the resonance frequencies, we fit a continuum model to the measured frequencies. We thereby retrieve the elastic constants of our (processed) material, namely the built-in stress $\sigma_0 = 830\,{\rm MPa}$ and Young's modulus $E=160\,{\rm GPa}$. From the displacement-induced, mode-dependent strain distribution, we calculate (except for an overall scaling) the mechanical quality factors. Calculated frequencies and quality factors are shown as hollow squares, the responses of the different harmonics of the same string are connected.
  }
\end{figure}

For each harmonic, we now are able to calculate the strain distribution within the resonator induced by the displacement $u[x]$ and exemplarily shown in Fig.\,1c. The measured dissipation is closely connected to this induced strain $\epsilon[x,z,t] = \epsilon[x,z] \exp[i 2 \pi f t]$. As in the case of a Zener model\,\cite{Cleland2003} we now assume that the displacement-induced strain and the accompanying oscillating stress $\sigma[x,z,t] = \sigma[x,z] \exp[i 2 \pi f t]$ are not perfectly in phase; this can be expressed by a Young's modulus $E = E_1 + i E_2$ having an imaginary part. The relation reads again $\sigma[x,z] = (E_1+i E_2)\epsilon[x,z]$. During one cycle of oscillation $T = 1/f$, a small volume $\delta V$ of length $s$ and cross section $A$ thereby dissipates the energy $\Delta U_{\delta V} = A s \pi E_2 \epsilon^2$. The total loss is obtained by integrating over the volume of the resonator.

\begin{equation}
\Delta U_{\rm n,m} = \int_V dV \Delta U_{\delta V} = \pi E_2 \int_V dV \epsilon_{\rm n,m}[x,z]^2
\label{eq_en_loss}
\end{equation}

The strain variation and its accompanying energy loss can be separated into contributions arising from overall elongation of the beam and its local bending. It turns out that here the former is negligible, despite the fact that the elastic energy is dominated by the elongation of the string, as discussed below. To very high accuracy we obtain for the dissipated energy $\Delta U_{\rm n,m} \approx \pi/12 E_2 w h^3 \int_l dx (\partial^2/(\partial x)^2u_{\rm n,m})^2$. A more rigorous derivation can be found in the Supplementary Information.\\
The total energy depends on the spatial mode (through $\epsilon_{\rm n,m}$, see exemplary Fig.\,1c) and therefore strongly differs for the various resonances. To obtain the quality factor, one has to calculate the stored energy e. g. by integrating the kinetic energy $U_{\rm n,m} = \int_l dx A \rho (2 \pi f_{\rm n,m})^2 u_{\rm n,m}[x]^2$. The overall mechanical quality factor is $Q= 2\pi U_{\rm n,m}/\Delta U_{\rm n,m}$. A more detailed derivation can be found in the Supplementary Information.

Assuming that the unknown value of the imaginary part $E_2$ of the elastic modulus is independent of resonator length and harmonic mode, we are left with one fit parameter $E_2$ to reproduce all measured quality factors and find excellent agreement (Fig.\,2, hollow squares). We therefore successfully model the damping of our nanoresonators by postulating a frequency independent mechanism caused by local strain variation. Allowing $E_2$ to depend on frequency, the accordance gets even better, as discussed in detail in the Supplement.

We now discuss the possible implications of our findings, considering at first the cause of the high quality factors in overall pre-stressed resonators and then the compatibility of our model with different microscopic damping mechanisms. In a relaxed beam, the elastic energy is stored in the flexural deformation and becomes for a small test volume $U_{\delta V}=1/2 A s E_1 \epsilon^2$. In the framework of a Zener model, as employed here, this result is proportional to the energy loss (see eq.\,\ref{eq_en_loss}) and thus yields a frequency-independent quality factor $Q = E_1/E_2$ for the unstressed beam. In accordance with this finding, Ref.\,\cite{Verbridge2006} reports a much weaker dependence of quality factor on resonance frequency, in strong contrast with the behavior of their stressed beams.

Similar as in the damping model, the total stored elastic energy in a beam can be very accurately separated into a part connected to the bending and a part coming from the overall elongation. The latter is proportional to the pre-stress $\sigma_0$ and vanishes for relaxed beams, refer to the Supplement for details. Assuming a constant $E = E_1 +i E_2$, Fig.\,3 displays the calculation of the elastic energy and the quality factor for the fundamental mode of our longest ($l=35\,\micro {\rm m}$) beam as a function of overall built in stress $\sigma_0$. The total elastic energy is increasingly dominated by the displacement-induced elongation $U_{\rm elong}=1/2 \sigma_0 w h \int_l dx (\partial/(\partial x)u[x])^2$. In contrast the bending energy $U_{\rm bend}=1/24 E_1 w h^3 \int_l dx(\partial^2/(\partial x)^2 u[x])^2$, which in our model is proportional to the energy loss, is found to increase much slower with $\sigma_0$. Thus one expects $Q$ to increase with $\sigma_0$, a finding already discussed by Schmid and Hierold for micromechanical beams\,\cite{Schmid2008}. However, their model assumes the simplified mode profile of a stretched string and can not explain the larger quality factors of higher harmonics when compared to a fundamental resonance of same frequency. Including beam stiffness, our model can fully explain the dependence of frequency and damping on length and mode index, as reflected in Fig.\,2. It also explains the initially surprising finding\,\cite{Southworth2009} that amorphous silicon nitride resonators exhibit high quality factors when stretched whereas having Q-factors in the relaxed state that reflect the typical magnitude of internal friction found to be rather universal in glassy materials\,\cite{Pohl2002}. More generally we conclude that the increase in mechanical quality factors with increasing tensile stress is not bound to any specific material.

Since the resonance frequency is typically easier to access in an experiment, we plot the quality factor vs. corresponding resonance frequency in Fig.\,3b; with both numbers being a function of stress. The resulting relation of quality factor on resonance frequency is (except for very low stress) almost linear; experimental results by another group can be seen to agree well with this finding\,\cite{Verbridge2007}. In addition, we show in the Supplement that although the energy loss per oscillation increases with applied stress, the linewidth of the mechanical resonance decreases.

\begin{figure}
  \includegraphics{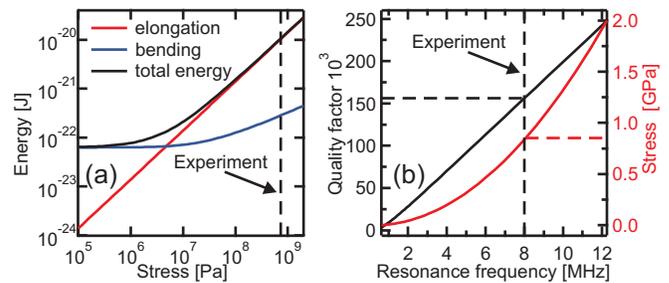}
  \caption{\label{fig3} {\bf Elastic Energy and Mechanical Quality Factor of the Beam in Dependence of Stress}
  {\bf a} The elastic energies of the fundamental mode of the beam with $l=35\micro$m are displayed vs. applied overall stress separated into the contributions resulting from the overall elongation and the local bending. The dashed line marks the strain of the experimentally studied resonator $\sigma_0 \approx 830$\,MPa, there the elongation term dominates noticeably. {\bf b} Quality factor and frequency are calculated for varying stress $\sigma_0$. In order to compare the calculation with other published results quality factor and stress are displayed vs. resulting frequency.
  }
\end{figure}

We will now consider the intrinsic physical mechanisms that could possibly contribute to the observed damping. As explained in greater detail in the Supplement, we can safely neglect clamping losses\,\cite{Hao2003,Wilson-Rae2008}, thermoelastic damping\,\cite{Lifshitz2000,Kiselev2008} and Akhiezer damping\,\cite{Akhieser1939,Kiselev2008} since they all predict damping constants significantly smaller than the ones observed.

Instead, we would like to discuss the influence of localized defect states. Similar to the Akhiezer effect, it is assumed that the energy spectra of defects are modulated by strain\,\cite{Jackle1972}, which thereby drives the occupancy out of thermal equilibrium. In Ref.\,\cite{Jackle1972}, this effect is calculated for a broad spectrum of two-level systems. The energy difference of two uncoupled levels as well as their separating tunnel barrier height are assumed to be uniformly distributed, leading to a broad yet not flat distribution of relaxation rates. In the high temperature limit the thus derived energy loss per oscillation becomes frequency-independent as assumed in our model. In addition, published quality factors of relaxed silicon nitride nano resonators\,\cite{Southworth2009} cooled down to liquid helium temperature display quality factors that are well within the typical range of amorphous bulk materials\,\cite{Pohl2002}.
Moreover, on a different sample chip we measured a set of resonators showing quality factors that are uniformly decreased by a factor of approximately 1.4 compared to the data presented in Fig.\,2. Their response can still be quantitatively modeled using now an increased imaginary part of Young's modulus $E_2$. We attribute this reduction to a non-optimized RIE-etch step, leading to an increased density of defect states. The corresponding data is presented in the Supplementary Information.
These three arguments clearly favor the concept of damping via defect states as the dominant mechanism.

We wish to point out that such a model calculation based on two-level systems cannot be rigourously applied at elevated temperatures, as the concept of two-level systems should be replaced by local excitable systems. However, it seems plausible that such a system still exhibits a broad range of relaxation rates, crucial to explain frequency-independent damping.  In contrast, a mechanism with discrete relaxation rates will exhibit damping maxima whenever the oscillation frequency matches the relaxation rate\,\cite{Lifshitz2000,Kiselev2008,Cleland2003}.
We further notice that in our experiments beams with larger widths exhibit slightly higher quality factors. This indicates an increased defect density near the surface being either intrinsic or caused by the micro-fabrication\,\cite{Yang2002} (RIE etch). For a fixed cross-section however, the applicability of our model is not affected.

In conclusion, we systematically studied the transverse mode frequencies and quality factors of prestressed SiN nanoscale beams. Implementing continuum theory, we reproduce the measured frequencies varying with beam length and mode index over an order of magnitude. Assuming that damping is caused by local strain variations induced by the oscillation, independent of frequency, enables us to calculate the observed quality factors with a single interaction strength as free parameter. We thus identify the unusually high quality factors of pre-stressed beams as being primarily caused by the increased elastic energy rather than a decrease in damping rate. Several possible damping mechanisms are discussed; because of the observed nearly frequency-independence of the damping parameter $E_2$, we attribute the mechanism to interaction of the strain with local defects of not yet identified origin. One therefore expects that defect-free resonators exhibit even larger quality factors, as recently demonstrated for ultra-clean carbon nanotubes\,\cite{Huttel2009}.

Financial support by the Deutsche Forschungsgemeinschaft via project Ko 416/18, the German Excellence Initiative via the Nanosystems Initiative Munich (NIM) and LMUexcellent as well as the Future and Emerging Technologies programme of the European Commission, under the FET-Open project QNEMS (233992) is gratefully acknowledged.

We would like to thank Florian Marquardt and Ignacio Wilson-Rae for stimulating discussions.

\newpage
\onecolumngrid

\begin{flushright}
\large \textsf{SUPPLEMENTARY INFORMATION} \rule{\linewidth}{0.5mm}
\end{flushright}

\section{Damping Model}

In a Zener model, an oscillating strain $\epsilon(t) = \Re[\epsilon[\omega] \exp[i \omega t]]$ and its accompanying stress $\sigma[t]= \Re[\sigma[\omega] \exp[i \omega t]]$ are out-of phase, described by a frequency-dependent, complex elastic modulus $\sigma(\omega) = E[\omega]\epsilon[\omega] = (E_1[\omega] + i E_2[\omega])\epsilon[\omega]$. This leads to an energy loss per oscillation in a test volume $\delta V = \delta A \cdot \delta s$ of cross-section $\delta A$ and length $\delta s$.
\begin{equation}
\Delta U_{\delta V} = \int_T dt  \underbrace{E A \epsilon[t]}_{\rm force} \cdot \underbrace{\frac{\partial}{\partial t} \left(s \epsilon[t] \right)}_{\rm velocity} = \pi \delta A \delta s E_2 \epsilon^2
\label{eq_en_loss}
\end{equation}
We now employ this model for our case, namely a pre-stressed, rectangular beam of length $l$, width $w$ and height $h$, corresponding here to the x,y,z-direction, respectively. The origin of the coordinate system is centered in the beam. The resonator performs oscillations in the z-direction and, as we consider a continuum elastic model, there will be no dependence on the y-direction. For a beam of high aspect ratio $l \gg h$ and small oscillation amplitude, the displacement of the $m$-th mode can be approximately written $u_m[x,y,z] = u_m[x]$. During oscillation, a small test volume within the beam undergoes oscillating strain $\epsilon_m[x,z,t]$.\\
This strain arises because of the bending of the beam as well as its elongation as it is displaced. The stress caused by the overall elongation is quadratic in displacement, therefore it occurs at twice the oscillating frequency.
\begin{align}
\epsilon_m[x,z,t] &= \underbrace{\frac{1}{2} \left(\frac{\partial}{\partial x}u_m[x] \Re[\exp[i \omega t]] \right)^2 }_{\rm elongation}+\underbrace{z \frac{\partial^2}{\partial x^2}u_m[x] \Re[\exp[i \omega t]]}_{\rm bending} \nonumber\\
&=
\frac{1}{2} \left(\frac{\partial}{\partial x}u_m[x] \right)^2 \frac{1}{2}\left(1+ \Re[\exp[2 i \omega t]] \right) +z \frac{\partial^2}{\partial x^2}u_m[x] \Re[\exp[i \omega t]]
\label{eq_strain}
\end{align}

Inserting this into eq.\,\ref{eq_en_loss} and integrating over the cross-section $w \cdot h$, the accompanying energy losses can be seen to separate into elongation and displacement caused terms.
\begin{equation}
\Delta U_{w \cdot h} = \pi s E_2[2 \omega] \frac{w h}{8} \left(\frac{\partial}{\partial x}u_m[x] \right)^4 +\pi s E_2[\omega] \frac{w h^3}{12} \left( \frac{\partial^2}{\partial x^2}u_m[x]\right)^2
\label{eq_loss_cross}
\end{equation}
Integrating over the length yields the total energy loss of a particular mode $\Delta U = \int_{-l/2}^{l/2} dx \Delta U_{w \cdot h}$. In the case that $E_2$ is only weakly frequency-dependent, it turns out that for our geometries the elongation term is several orders of magnitude ($10^5-10^7$) smaller than the term arising from the bending. The energy loss therefore may be simplified and writes
\begin{equation}
\Delta U \approx \Delta U_{\rm bending} = \pi E_2 \frac{wh^3}{12} \int_{-l/2}^{l/2} dx \left(\frac{\partial^2}{\partial x^2} u_m[x] \right)^2
\end{equation}

\section{Elastic Energy of a Pre-Stressed Beam}

A volume $\delta V$ subject to a longitudinal pre-stress $\sigma_0$ stores the energy $U_{\delta V}$ when strained; $E_1$ is assumed to be frequency independent in the experimental range (5-100\,MHz)
\begin{equation}
U_{\delta V} = s A \left(\sigma_0 \epsilon + \frac{1}{2} E_1 \epsilon^2 \right)
\end{equation}
To apply this formula to the case of an oscillating pre-stressed beam, we insert eq.\,\ref{eq_strain}$|_{t=0}$ (maximum displacement) and integrate over the cross-section to obtain
\begin{equation}
U_{w \cdot h} =
\frac{1}{2} E_1 \left(\frac{1}{4} w h \left(\frac{\partial}{\partial x}u_m[x]\right)^4 +
\frac{1}{12} w h^3 \left(\frac{\partial^2}{\partial x^2} u_m[x] \right)^2
\right)+
\frac{1}{2} s w h \sigma_0 \left(\frac{\partial}{\partial x} u_m[x]\right)^2
\end{equation}
Analog to eq.\,\ref{eq_loss_cross} we can omit the first term in the brackets; integrating over the length yields
\begin{equation}
U \approx \int_{-l/2}^{l/2} dx \Big( \underbrace{\frac{1}{2} w h \sigma_0 \left(\frac{\partial}{\partial x} u_m[x]\right)^2}_{\rm elongation} +
\underbrace{ \frac{1}{24} E_1 w h^3 \left(\frac{\partial^2}{\partial x^2} u_m[x] \right)^2
}_{\rm bending} \Big)
\end{equation}
We can therefore divide the total energy into parts arising from the elongation and the bending of the beam. Depending on the magnitude of the pre-stress, either of the two energies can dominate as seen in Fig.\,3a of the main text.
We have checked that the kinetic energy $U_{\rm kin} = 1/2 \rho (\omega_m)^2 \int_{l/2}^{l/2} dx (u_m[x])^2$; ($\omega_m/(2\pi),\rho$: resonance frequency, material density, respectively) equals the total elastic energy, as expected.

\section{Frequency-dependent Loss Modulus}

There is no obvious reason that the imaginary part of Young's modulus $E_2$ should be completely frequency-independent. We therefore assume that $E_2$ obeys a (weak) power-law and chose the ansatz:
\begin{equation}
E_2[f] = E_2 (f/f_0)^a
\end{equation}
Fitting our data with the thus extended theory, we achieve a very precise agreement of measured and calculated quality factors, as seen in Fig.\,S1. The resulting exponent is $a=0.075$; $E_2$ varies therefore by 20\% when $f$ changes by one order of magnitude.\\
\begin{figure}[h]
\begin{center}
  \includegraphics{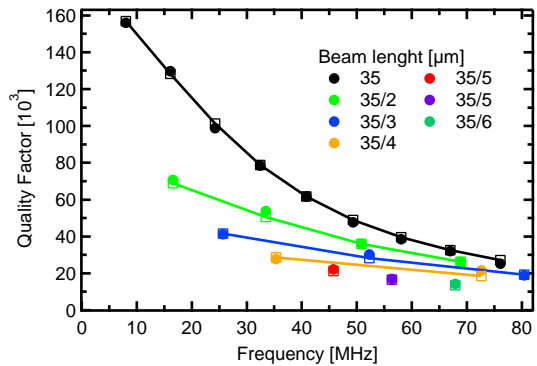}
\end{center}
  \caption{\label{fig2} \emph{Resonance frequencies and quality factors of the resonators}
 {\bf a} Measured quality factor and resonance frequency of several harmonics of beams with different lengths (color-coded) are displayed as filled circles (same data as in Fig.\,2 of the main text). The resonance frequencies are reproduced by a continuum model; we calculate the quality factors using a model based on the strain caused by the displacement. In contrast to Fig.\,2 of the main text and Fig.\,S2 we here allow $E_2$ to be (weakly) frequency-dependent.
}
\end{figure}

\section{Linewidth of the Mechanical Resonance}

The elastic energy of a harmonic oscillator is given by $U = 1/2 m_{\rm eff} \omega_0^2 x_0^2$ with $m_{\rm eff}, \omega_0, x_0$ being effective mass, resonance frequency and displacement, respectively. If we assume the effective mass to be energy-independent, it applies $\omega_0 \propto \sqrt U$.
Recalling the definition of the quality factor  $Q = 2\pi U/\Delta U \propto U$, one obtains for the for the Full Width at Half Max (FWHM) of the resonance
\begin{equation}
\Delta \omega = \frac{\omega_0}{Q}  \propto \frac{\sqrt U}{U/\Delta U} = \frac{\Delta U}{\sqrt{U}}
\end{equation}

As in the main text, the energy depends on the applied overall tensile stress. Figure\,\ref{fig3} shows a numerical calculation of the resulting linewidth vs. applied stress; one can see that increase in energy loss per oscillation is dominated by the increase in energy, resulting in a decreased linewidth. The exact effective mass is included in this calculation; as it changes by less than 20\%, the above assumption is justified.

\begin{figure}[h]
\begin{center}
  \includegraphics{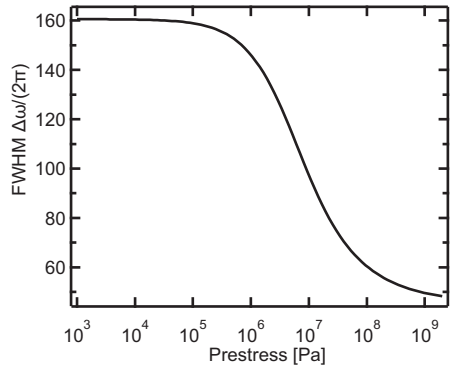}
\end{center}
  \caption{\label{fig3} \emph{Linewidth of the mechanical resonance}
  The calculated linewidths (FWHM) for the fundamental mode of the beam with $l=35\micro$m are displayed vs. applied overall stress.
}
\end{figure}

\section{Microscopic Damping Mechanisms}

We start with clamping losses as discussed, e.g., in ref.\,\cite{Hao2003,Wilson-Rae2008}, i. e. the radiation of acoustic waves into the bulk caused by inertial forces exerted by the oscillating beam. With a sound velocity in silicon of $v_{\rm Si} \approx 8$\,km/s, the wavelength of the acoustic waves radiated at a frequency of 10\,MHz from the clamps into the bulk will be greater than 500\,$\micro$m, and thus substantially larger than the length of our resonators. Considering each clamping point as a source of an identical wave propagating into the substrate, one would expect that mostly constructive/destructive interference would occur for in-/out-of-phase shear forces exerted by the clamping points, respectively. With clamping losses being important, one would therefore expect that spatially asymmetric modes with no moving center of mass exhibit significant higher quality factors than symmetric ones\,\cite{Ignacio}. Another way to illuminate this difference is that symmetric modes give rise to a net force on the substrate, whereas asymmetric modes yield a torque. Since the measurement (Fig. 2) does not display such an alternating behavior of the quality factors with mode index $m$ (best seen for the longest beam), clamping losses are likely to be of minor importance.

The next damping mechanism we consider are phonon-assisted losses within the beam. At elevated temperatures, at least two effects arise, the first being thermoelastic damping: because of the oscillatory bending, the beam is compressed and stretched at opposite sides. Since such volume changes are accompanied by work, the local temperature in the beam will deviate from the mean. For large aspect ratios as in our case, the most prominent gradient is in the z direction. The resulting thermal flow leads to mechanical dissipation. We extend existing model calculations\,\cite{Lifshitz2000} to include the tensile stress of our beams.
Using relevant macroscopic material parameters such as thermal conductivity, expansion coefficient and heat capacity we derive Q-values that are typically three to four orders of magnitudes larger than found in the experiment. Therefore, heat flow can be safely neglected as the dominant damping mechanism. In addition, the calculated thermal relaxation rate corresponds to approximately 2\,GHz, so the experiment is in the so-called adiabatic regime. Consequently, one would expect the energy loss to be proportional to the oscillation frequency, in contrast to the assumption of a frequency independent $E_2$ and our experimental findings.

Another local phonon-based damping effect is the Akhiezer-effect\,\cite{Akhieser1939}; it is a consequence of the fact that phonon frequencies are modulated by strain, parameterized by the Gr\"uneisen tensor. If different phonon modes (characterized by their wave vector and phonon branch) are affected differently, the occupancy of each mode corresponds to a different temperature. This imbalance relaxes towards a local equilibrium temperature, giving rise to mechanical damping. In a model calculation applying this concept to the oscillatory motion of nanobeams\,\cite{Kiselev2008}, the authors find in the case of large aspect ratios length/height that the thermal heat flow responsible for thermoelastic damping dominates the energy loss by the Akhiezer effect. We thus can safely assume this mechanism to be also negligible in our experiment.

\section{Reduced Quality Factor}

We fabricated a set of resonators, shown in Fig.\,S1a, that showed lower quality factors than the ones presented in the main text (Fig.\,2); we attribute this reduction to a non-optimized RIE-etch step. As in the main article, it is possible to reproduce the quality factors using a single fit parameter, namely the imaginary part of Young's modulus $E_2$. The ratio of the two sets of quality factors is displayed in Fig.\,S1 b and can be seen to be around 1.4 with no obvious dependence on resonance frequency, mode index or length. A non-optimized etch step causes additional surface roughness and the addition of impurities, thereby increasing the density of defect states. As there is no obvious reason why another damping mechanism should be thereby influenced, we interpret this as another strong indication that the dominant microscopic damping mechanism is caused by localized defect states.
\begin{figure}[h]
\begin{center}
  \includegraphics{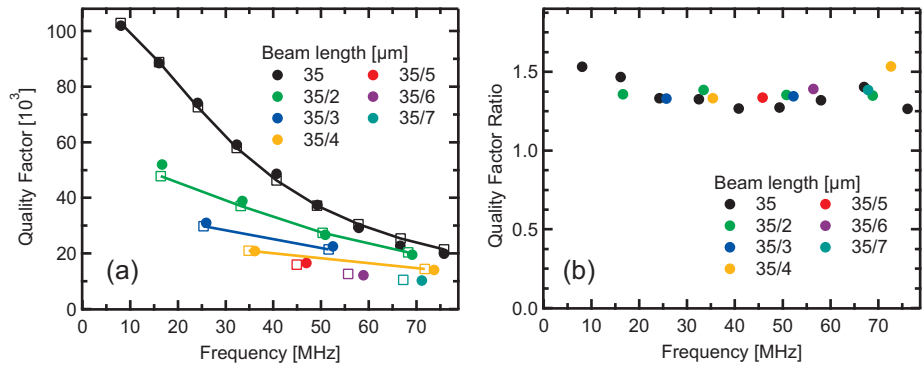}
\end{center}
  \caption{\label{fig3} \emph{Comparison of the resonance frequencies and quality factors of the sets of resonators}
 {\bf a} Measured quality factor and resonance frequency of several harmonics of beams with different lengths (color-coded) are displayed as filled circles. The resonance frequencies are reproduced by a continuum model; a model based on the strain caused by the displacement allows us to calculate the quality factors, shown as hollow squares. The uniform reduction of the Q-factors is attributed to an non-optimized RIE-etch. {\bf b} The ratio of the quality factors of the two sets resonators (Fig.\,2 main article and Fig.\,S2a) are displayed vs. frequency, being approximately constant.
}
\end{figure}

\end{document}